\documentclass[sigconf, nonacm]{acmart}

\usepackage{listings}

\newcommand\vldbdoi{XX.XX/XXX.XX}
\newcommand\vldbpages{XXX-XXX}
\newcommand\vldbvolume{14}
\newcommand\vldbissue{1}
\newcommand\vldbyear{2020}
\newcommand\vldbauthors{\authors}
\newcommand\vldbtitle{\shorttitle} 
\newcommand\vldbavailabilityurl{https://github.com/matthijsr/til-vhdl}
\newcommand\vldbpagestyle{plain}

\begin{document}

\title{An Intermediate Representation for Composable Typed Streaming Dataflow Designs}

\author{Matthijs A. Reukers}
\orcid{0000-0002-1612-1187}

\author{Yongding Tian}
\orcid{0000-0002-4515-4009}

\author{Zaid Al-Ars}
\orcid{0000-0001-7670-8572}

\affiliation{%
  \institution{Delft University of Technology}
  \streetaddress{Mekelweg 5}
  \city{Delft}
  \country{The Netherlands}
  \postcode{2600 AA}}

\email{M.A.Reukers@student.tudelft.nl}
\email{{Y.Tian-3,Z.Al-Ars}@tudelft.nl}

\author{Peter Hofstee}
\orcid{0000-0002-1825-0097}
\affiliation{%
  \institution{IBM Systems}
  \city{Austin}
  \state{TX}
  \country{USA}
}
\email{H.P.Hofstee@tudelft.nl}

\author{Matthijs Brobbel}
\orcid{0000-0001-6286-6510}

\author{Johan Peltenburg}
\orcid{0000-0002-7043-7131 }

\author{Jeroen van Straten}
\orcid{0000-0002-5610-2511}

\affiliation{%
  \institution{Voltron Data}
  \city{Mountain View}
  \state{CA}
  \country{USA}}

\email{{matthijs,johan,jeroen}@voltrondata.com}

\renewcommand{\shortauthors}{Reukers, et al.}

\begin{abstract}
  Tydi is an open specification for streaming dataflow designs in digital circuits, allowing designers to express how composite and variable-length data structures are transferred over streams using clear, data-centric types. These data types are extensively used in many application domains, such as big data and SQL applications. This way, Tydi provides a higher-level method for defining interfaces between components as opposed to existing bit- and byte-based interface specifications. In this paper, we introduce an open-source intermediate representation (IR) which allows for the declaration of Tydi's types. The IR enables creating and connecting components with Tydi Streams as interfaces, called Streamlets. It also lets backends for synthesis and simulation retain high-level information, such as documentation. Types and Streamlets can be easily reused between multiple projects, and Tydi’s streams and type hierarchy can be used to define interface contracts, which aid collaboration when designing a larger system. The IR codifies the rules and properties established in the Tydi specification and serves to complement computation-oriented hardware design tools with a data-centric view on interfaces. To support different backends and targets, the IR is focused on expressing interfaces, and complements behavior described by hardware description languages and other IRs. Additionally, a testing syntax for the verification of inputs and outputs against abstract streams of data, and for substituting interdependent components, is presented which allows for the specification of behavior. To demonstrate this IR, we have created a grammar, parser, and query system, and paired these with a backend targeting VHDL.
  
  
  
\end{abstract}



\maketitle

\pagestyle{\vldbpagestyle}
\begingroup\small\noindent\raggedright\textbf{PVLDB Reference Format:}\\
\vldbauthors. \vldbtitle. PVLDB, \vldbvolume(\vldbissue): \vldbpages, \vldbyear.\\
\href{https://doi.org/\vldbdoi}{doi:\vldbdoi}
\endgroup
\begingroup
\renewcommand\thefootnote{}\footnote{\noindent
This work is licensed under the Creative Commons BY-NC-ND 4.0 International License. Visit \url{https://creativecommons.org/licenses/by-nc-nd/4.0/} to view a copy of this license. For any use beyond those covered by this license, obtain permission by emailing \href{mailto:info@vldb.org}{info@vldb.org}. Copyright is held by the owner/author(s). Publication rights licensed to the VLDB Endowment. \\
\raggedright Proceedings of the VLDB Endowment, Vol. \vldbvolume, No. \vldbissue\ %
ISSN 2150-8097. \\
\href{https://doi.org/\vldbdoi}{doi:\vldbdoi} \\
}\addtocounter{footnote}{-1}\endgroup

\ifdefempty{\vldbavailabilityurl}{}{
\vspace{.3cm}
\begingroup\small\noindent\raggedright\textbf{PVLDB Artifact Availability:}\\
The source code, data, and/or other artifacts have been made available at \url{\vldbavailabilityurl}.
\endgroup
}

\section{Introduction}\label{sec:introduction}

In order to transfer streaming data between components within digital circuits, designers have a choice to either design their own interfaces, or use general interface specifications such as Intel's Avalon-ST \cite{intelcorporationavalon2022} or Arm's AXI4-Stream \cite{armlimitedamba2010}. By using an interface specification, it is easier for other designers to connect components, as the signals and how they relate to data transfers are standardized. This can promote reuse, and is used by hardware design tools to provide IP (Intellectual Property) libraries and automate integration \cite{arnesenincreasing2010, jacomesurvey2001}.

The aforementioned specifications do not specify how data structures are represented, however, and as a result designers must still design, document and share these representations. Additionally, the IP integration tools are proprietary, reducing the simplicity of integrating such IPs outside of these specific tools. Addressing the first issue, Peltenburg et al. proposed Tydi (Typed dataflow interface) \cite{peltenburgtydi2020}, an open specification which allows designers to explicitly define the data which is being transferred by providing a type system for composite and variable-length data structures, in addition to defining how data elements are organized in transfers and the requirements on transfers. This paper introduces a method to address the second issue, by utilizing the Tydi specification as part of an IR (intermediate representation) for defining interfaces and connecting components.

The goal of the IR is not to serve as a complete hardware description language, but to provide a simple and robust way to declare Tydi's types, define interfaces and connect components which adhere to the Tydi specification, serving as part of a toolchain in order to integrate and reuse components within and across projects. To this end, the IR is not capable of directly implementing behavior, but should instead be combined with transaction-level verification to \textit{specify} intended behavior.

\section{Related Work}\label{sec:related}

The Tydi specification and type system was introduced by Peltenburg et al. \cite{peltenburgtydi2020} and defines an abstract way to describe data structures transferred over hardware streams. Tydi promises to reduce the design effort of creating hardware for streaming dataflow computing, by providing clear and intuitive ways to map composite, variable-length data structures onto a hardware streaming protocol. An open-source repository and documentation \cite{vanstratenintroduction2021} expanding on the specification and providing example code for mapping Tydi's streams onto VHDL component ports is now available.

\textbf{Stream processing} There exist many projects to introduce languages and frameworks for streaming data processing \cite{thiesstreamit2002, hormatioptimus2008, auerbachlime2010, saxapache2018, thomasfleet2020, octoray}, as stream processing has been actively researched for over 20 years and has undergone many changes, with software paradigms and hardware acceleration being worked on in parallel \cite{fragkoulissurvey2020, isahsurvey2019, bigdata_joost}.

\textbf{Design tools} At the same time, there are multiple ongoing efforts to improve the tools used for designing such hardware accelerators, in the form of new hardware description languages \cite{bachrachchisel2012, izraelevitzreusability2017}, intermediate representations \cite{schuikillhd2020} and compilers \cite{llvmcommunitycirct2022}, high-level synthesis based on software programming languages \cite{nanesurvey2016}, and more general program representations for heterogeneous systems \cite{plavecstream2010, kotsifakouhpvm2018}.

\textbf{Interfaces} While improved languages and frameworks can speed up design and improve reusability \cite{izraelevitzreusability2017,fletcher, fletcher_internals}, interface standards such as \cite{armlimitedamba2010, armlimitedintroduction2021, intelcorporationavalon2022, pontesscaffi2007} ensure that IPs can be made compatible between projects. Additionally, design tools provide catalogs of IPs and methods to combine and integrate them into larger designs \cite{advancedmicrodevicesinc.intellectual2022, intelcorporationintroduction2021}.


\textbf{Tydi} Working towards a full toolchain, there is a separate project for a front-end and language for expressing behavior, called Tydi-lang~\cite{tydilang}.


\section{Motivation}\label{sec:background}



While much research is focused on developing and accelerating algorithms for streaming data in both hardware \cite{nowatzkistreamdataflow2017, plavecstream2010} and software \cite{isahsurvey2019}, many designs for low-level hardware still have to transfer streams over interfaces which are either custom or based on generic, bit- and/or byte-oriented specifications such as AXI4-Stream \cite{armlimitedamba2010} and Avalon-ST \cite{intelcorporationavalon2022}. As a result, higher-level information about data structures and how streams of data are organized over transfers must be devised and implemented by designers, and are not reflected by the declaration of the interface in a traditional HDL.


Some of this design effort can be alleviated through the use of high-level synthesis: tools such as Vivado HLS can be employed to leverage C, C++ or SystemC combined with IP-blocks using \textit{ap\_fifo} or AXI4-Stream to handle data streams \cite{advancedmicrodevicesinc.interfaces2022}, while synthesizing compilers such as Optimus \cite{hormatioptimus2008} have been developed in the past to leverage StreamIt \cite{thiesstreamit2002}, a language specifically for streaming applications. At the same time, many researchers are working on improved hardware description languages and IRs, such as Chisel \cite{bachrachchisel2012}, FIRRTL \cite{izraelevitzreusability2017} and LLHD \cite{schuikillhd2020}.

These are not suitable replacements for a higher-level interface specification, however: HLS tools either obfuscate the interfaces between low-level hardware and/or use proprietary IRs and tools to connect components, making reuse more difficult. While the HDLs and IRs mentioned are aimed at more general hardware designs, so still require custom interfaces for streaming data transfers. As such, we propose a free, open-source IR for defining high-level streaming dataflow interfaces mapped onto hardware and for connecting these interfaces. This complements existing HDLs and IRs which describe behavior, and enabling components designed in higher-level front-end languages for HLS to propagate more type information to the resulting interfaces.

\section{Type Declarations and Interface Design}\label{sec:types_interfaces}
\subsection{Type Hierarchy and Complex Data Structures}\label{subsec:types}


The Tydi specification \cite{vanstratenintroduction2021} defines five \textit{logical types}: the stream-manipulating Stream type, and the element-manipulating Null, Bits, Group and Union types. The intermediate representation features the ability to compose these, and declare them with a unique identifier in a namespace.

In short, the Null type is for transfers of one-valued data (its only valid value is null), Bits(N) represents a data signal of N bits, while the Group and Union types contain \textit{fields} consisting of a unique name and a logical type. Groups and Unions are distinct in that Groups are composites of multiple types, where each field is set at the same time, while Unions are exclusive disjunctions of types, where only one field can be active at a time, to be selected with a \textit{tag} signal. Finally, the Stream type represents a new physical stream carrying these types.

The element-manipulating types alone can represent many data structures, for example: Bits(N) can be used to transfer primitive data types such as numbers, booleans and characters, a Union of Null and another type can indicate optional data, and Groups can be used to represent records of data. However, as Streams are also logical types, Groups, Unions and Streams themselves can carry further nested logical Streams, each with their own data and properties.

The Stream type adds a further layer of flexibility to these types. It does not only represent the physical stream and signals carrying the element-manipulating types, but also features properties for further describing data structures. Notably, Streams have a \textit{dimensionality} property, which indicates whether the data being transferred is part of a sequence. In hardware, this is translated to a ``last'' signal; when this signal is driven high, it indicates that the data being transferred is the last element in a sequence, and a Stream with a higher dimensionality will have multiple last signals, to indicate nested sequences.

In addition to dimensionality, Streams have properties for describing how transfers should be organized in space and time, as follows:

\begin{itemize}
    \item \textit{Throughput} is a positive, rational number indicating how many elements are expected to be transferred per individual handshake, or relative to its parent Stream. The number of element lanes is \textit{throughput} rounded up to a natural number.
    \item \textit{Direction} indicates whether a Stream flows in the same direction as its parent, or in reverse. As an example, a Group can have both a ``Forward'' and ``Reverse'' Stream, for indicating that interdependent data is transferred between the sink and source, such as a memory address and the data retrieved from that address.
    \item \textit{Synchronicity} refers to how strong the relation between a child Stream and its parents are with regards to dimensional information. ``Sync'' indicates that for each element transferred on the parent, the child has a matching transfer, while ``Desync'' indicates that the child may have transfers of arbitrary size. Both options also have a ``Flat'' variant, which results in redundant last signals on the child being omitted.
    \item \textit{Complexity} is a number which encodes guarantees on how elements of a sequence are transferred. Overall, a lower complexity imposes more restrictions on a source, which conversely results in a higher complexity making it more difficult to implement a sink. As an example, a complexity of $\leq2$ requires that elements of an inner sequence are transferred over consecutive cycles by a source, while higher complexities allow it to stall independently from the sink. The specification currently defines 8 levels of complexity \cite{vanstratenphysical2021}.
    \item A \textit{keep} property can be used to ensure a \textit{logical} Stream is synthesized into physical signals, as nested Streams may otherwise be combined into a single physical stream.
\end{itemize}

\begin{figure}[h]
  \centering
\includegraphics[width=\linewidth]{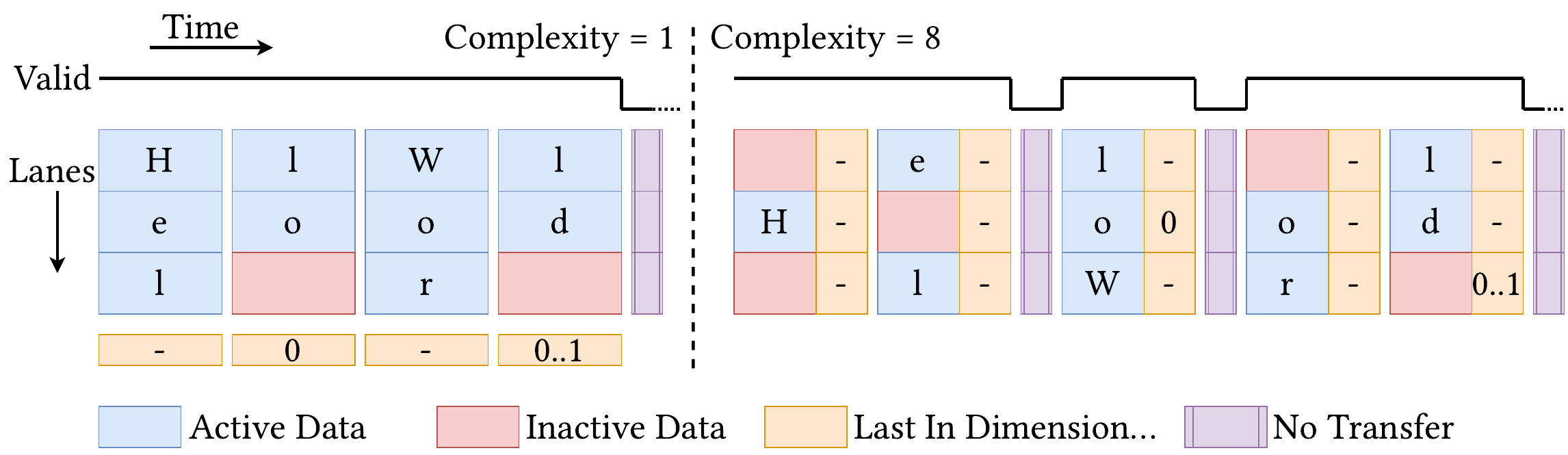}
  \caption{Streams determine which signals are used and valid to organize elements in transfers, and how transfers are organized over time.}
  \label{fig:streamproperties}
\end{figure}

Figure \ref{fig:streamproperties} illustrates how a higher \textit{complexity} allows for transfers to be organized differently. When transferring \textit{[[H, e, l, l, o], [W, o, r, l, d]]}, at \textit{complexity} $=1$ all elements must be aligned to the first lane, \textit{last} data is asserted per transfer, and all data must be transferred over consecutive cycles and lanes. At \textit{complexity} $=8$, there are no requirements for how elements are aligned, transfers may be postponed (asserting \textit{valid} low), and \textit{last} data is asserted per lane, and may be postponed (using an inactive lane to assert \textit{last} for a previous lane or transfer).

Finally, in the event these properties are insufficient for a use-case, Streams can also have a \textit{user} signal carrying an element-manipulating type. This user signal can be used to provide additional information independent from transfers or clock cycles.




\subsection{Interfaces as Contracts}\label{subsec:interfaces}

\subsubsection{Communicating Intent}

As the previous section would suggest, Tydi's types can convey a significant amount of information; not just what data is transferred, but also how it is transferred, and how sequences of elements relate to one another. In effect, a sufficiently detailed Stream definition can be treated as a \textit{contract} between components (and in a sense, designers) on how a stream of data will be implemented.

The intermediate representation builds on this when declaring \textit{Interfaces}. In its simplest form, an Interface represents a collection of ports on a component (Streamlet), each of which carries a logical Stream either into or out of the component.

However, each Interface and its ports may also feature \textit{documentation}. Distinct from comments on a grammar, documentation is an actual property of a port or interface, and is expected to be implemented by a backend, typically by generating matching comments on the related output. Documentation being propagated from higher-level descriptions to the actual computation-oriented design tools that the IR complements is primarily useful when either implementing a component based on an interface template, or when trying to identify how physical signals relate to their abstract definition.

While Tydi's Streams assume a single clock and reset signal, which together make up their clock and reset domain, regardless of how many physical streams they are composed of, the ports of an Interface do not need to rely on the same clock and reset signals. Instead, an Interface may have one or more uniquely named \textit{domains} which represent a clock and reset signal, each of which is associated with one or more of the Interface's ports.

Subsequently, while the intermediate representation does not feature the ability to define a specific clock or how a reset signal should be handled, designers can use these domains to ensure multiple clock and reset signals are available on a component, and that ports which belong to different domains are not directly connected. In the event no domain is specified on the Interface, a default domain is instead created and assigned to all ports, as Tydi currently only defines Streams in the context of a clock.







\subsubsection{Notes on Interface Compatibility}\label{subsec:interface_compat}


The ports of Interfaces are compatible with one another when they have the same logical type, appropriate directions (for each physical stream, there is a source and matching sink), and the same clock domain.

Note that while types in the IR may be defined with identifiers, these identifiers are not a property of the logical type in question, and only exist within the namespace. This choice was made to restrict the IR to properties defined in the Tydi specification.

As a result, types with different names but otherwise identical properties are fully compatible; on an abstract level, this can be interpreted as a kind of implicit casting between types. Although in our evaluation with respect to readability of backend output, discussed in Section \ref{subsec:readability}, and in light of the potential added value of a stricter type system, we may reconsider this approach in the future. For instance, we may make identifiers an intrinsic property of types, and separately support type aliases for functionality similar to the current behavior.

However, while type identifiers are not currently relevant to compatibility, \textit{field} identifiers are an actual property of the Group and Union types. Hence, a \texttt{Group(a: Null)} is not compatible with a \texttt{Group(b: Null)}, regardless of whether they are physically identical.

Finally, while \textit{complexity} is a property of the Stream type, the Tydi specification does conditionally allow Streams with different complexities but otherwise identical properties to be connected. Specifically, a physical \textit{source} stream may be connected to a \textit{sink} if its complexity is equal to or lower than that of the sink. Note however that this applies to physical streams: logical Streams do not have a notion of sinks and sources, and may contain child Streams which flow in reverse directions, resulting in them containing both sink and source physical streams.

As such, the IR considers the Streams of ports incompatible when their complexity is not identical. While the process of connecting compatible physical streams can be optimistically automated to improve reuse, as discussed later in Section \ref{subsec:intrinsics}, designers should generally strive for a shared, normalized complexity between Streams.



\section{Component Composition and Implementation}\label{sec:component_implementation}

In addition to Interfaces, the IR introduces the ability to declare components, referred to as \textit{Streamlets}. These Streamlets consist of an Interface and optionally an Implementation. In effect, there are two different kinds of Implementation for a Streamlet: a \textit{structural} implementation, which can be used to combine instances of streamlets into a larger design, and a \textit{link} to an implementation of behavior in the target language or format.

Streamlets are the intended output of a project; Types, Interfaces and Implementations are not expected to be included in a backend's emissions unless they are part of a Streamlet, but can be shared between IR projects.

As Streamlets always have an Interface, they can be \textit{subsetted} to Interfaces, which can be used to express alternate implementations of the same component, e.g. when versioning a component or when substituting one for the purposes of testing as described in Section \ref{subsec:substitution}.

\subsection{Structural Composition}

As the goal of both Tydi and the IR is to improve compatibility and reuse of primitive components, the IR features the ability to connect Streamlets to one another. The IR refers to this as a \textit{Structural} implementation.

Structural implementations can contain \textit{instances} of Streamlets and connections between ports of Streamlets. Instances consist of a local name and a reference to a Streamlet declaration, the ports of their interfaces are assigned separately through connections.

Connections can be created between the ports of both Streamlet instances and the containing Streamlet which is being implemented, and require both ports to have identical types and clock domains (for the reasons described in Section \ref{subsec:interface_compat}). Connections are explicitly not ``assignments'', as the direction of a port is already known, and there is not necessarily one overall direction for a Stream type due to the possibility to define Streams which are \textit{Reversed} (such as when representing request and response streams). Hence, the \textit{source} and \textit{sink} between two ports of a connection is determined during lowering for each resulting Physical Stream.

By default, the IR requires that each port of each Streamlet is connected to exactly one other port. Leaving ports unconnected is against the Tydi specification, which requires that a default signal is driven for omitted signals \cite{vanstratenphysical2021}. While HDLs such as VHDL and Verilog support one-to-many and many-to-one connections at a signal-level, these are not allowed by the IR due to the fact that ports represent Streams with handshake signals, which would need to be combined.

While combining the \textit{ready} signals of multiple sinks could be achieved with simple logical \textit{and} expressions for a one-to-many connection, combining multiple transfers in a many-to-one connection has no clear, universally applicable solution. Even the aforementioned one-to-many implementation is not universal, as some designs may call for only one of the many to accept a transfer. Finally, as a connection does not necessarily have a single direction, a one-to-many connection between ports may well contain physical many-to-one transfers.






\subsection{Linked Implementations}

The intermediate representation intentionally omits expressions for implementing or simulating arbitrary behavior of components. Designing a language or set of expressions for functional hardware design and simulation is a difficult problem which is already being addressed by many researchers and organizations, elaborated in section \ref{sec:related}. Instead, ``behavioral implementations" in the IR exist only as \textit{links} to directories, which contain the relevant code in languages more suited for expressing behavior.

How these links are used is left up to the backend, though a simple use-case would be to create or copy a file in the target output language based on the Streamlet's name. As these are directories, multiple such files can exist side-by-side for different targets, and implementations do not need to be constrained to a single file; a linked directory could even be used to refer to a project or library consisting of multiple files, provided this exposes the Interface of the Streamlet being implemented.

\begin{figure}[h]
  \centering
  \includegraphics[width=\linewidth]{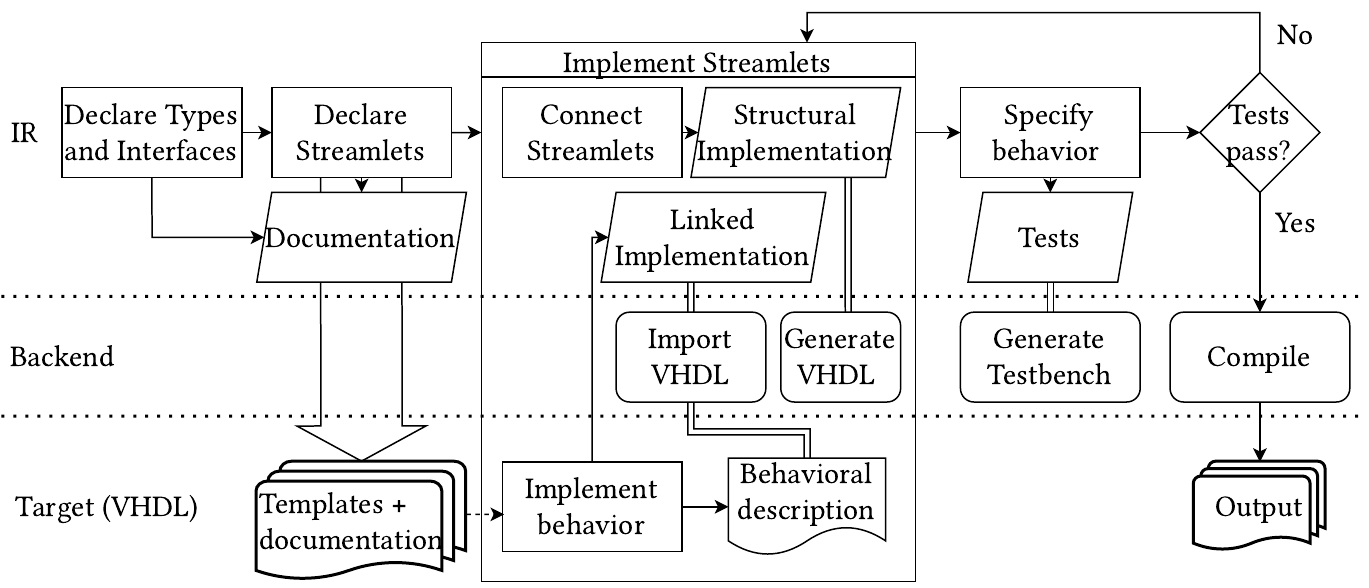}
  \caption{An example workflow, demonstrating how Streamlets are implemented using the IR, a suitable backend, and behavior defined in the target language.}
  \Description{A flowchart with three lanes separated by dotted lines. The top lane is called "IR", the middle lane is called "Backend" and the bottom lane is called "Target (VHDL)".}
  \label{fig:irworkflow}
\end{figure}

Figure \ref{fig:irworkflow} illustrates how linked implementations fit within a partial toolchain and workflow, consisting of Streamlets, structural implementations and tests defined in the IR, combined with behavior defined in a target language (VHDL, in this example) by a suitable backend. Not pictured are tools for simulating the testbenches produced by the backend, further passes on the output, nor any potential frontend language.

\subsection{Intrinsics}\label{subsec:intrinsics}

While the intermediate representation does not support expressing \textit{arbitrary} functionality, we do recognize a subset of functionality useful for implementing Tydi-based components and streaming dataflow designs in general. For general design, slices, buffers, and general-purpose stream manipulating components such as synchronizers are obvious candidates, while methods for optimistically connecting Streams with different complexities, or driving default or constant values to otherwise unconnected ports could help when reusing existing Streamlet designs.

Hence, we propose establishing a minimal, portable set of intrinsic functions, or \textit{intrinsics}, to be implemented by any backend. Specifically, intrinsics should only cover commonly used, simple functionality which cannot be implemented by a library of fixed component designs; as an example, slices are commonly used and simple in both their functionality and implementation, but a fixed library cannot address each possible interface design.

Another useful kind of intrinsic or backend feature would be one to improve readability and communicate intent in the target language. For example, Tydi's documentation mentions permitted alternative representations of interfaces \cite{vanstratenphysical2021}, which can leverage data types and arrays to improve readability. These alternative representations could be automatically generated for empty or template linked implementations by the backend, and wrapped in components using the conventional signals, clarifying the relation between signals and their logical type definitions for designers working in the target language.

A broader kind of ``intrinsic'' would be features such as generic properties of types, and generating loops. Both can be implemented as language features (either on the intermediate language, or a front-end) and evaluated without the backend's knowledge. However, by including these in the IR, backends for languages which adequately support these features could propagate them directly.



\section{Specification}\label{sec:specification}
While the intermediate representation lacks the ability to completely implement behavior, it can nonetheless allow for the specification of behavior through tests.

\subsection{High-level Assertions}\label{subsec:assertions}

As the IR is used to represent ports consisting of Streams carrying logical types, it is best suited for transaction-level verification. Inputs and outputs should be verified against abstract streams of data, upon which the IR combined with a backend will generate the necessary signaling behavior and assertions. This enables designers to verify the behavior of components and correctness of their interfaces without needing to concern themselves with the target language.

There are two key properties to consider when designing and generating tests for Interfaces:
\begin{enumerate}
    \item Ports of an Interface are not required to be interdependent or synchronized with one another.
    \item A port's Stream does not necessarily have a single direction, as child Streams can be \textit{Reversed}.
\end{enumerate}

To address these, we propose a testing grammar with the following properties: First, transaction verification on ports should be assumed to happen in parallel by default, rather than in the sequence assertions are declared. For example, implementing a Streamlet which adds two inputs could be represented as follows, assuming the output does not assert \textit{valid} until it has received and added two inputs:

\begin{lstlisting}[basicstyle=\ttfamily\scriptsize]
adder.out = ("10", "01", "11");
adder.in1 = ("01", "01", "10");
adder.in2 = ("01", "00", "01");
\end{lstlisting}

Where \textit{("10", "01", "11")} represents a series of \textit{Bits(2)} to be transferred over a Stream without dimensionality. This is to be transferred depending on throughput; e.g., one port could support two elements per transfer and require only two transfers, while another might only support one element per transfer and require three. In this proposed syntax, square brackets would be used to indicate dimensionality: \textit{[["1", "0"], ["0"]]}

Second, rather than explicit \textit{assign} and \textit{compare} methods, the IR should automatically determine whether physical streams are sinks or sources. The latter property means that something closer to mathematical equality is implemented; ``the transaction on \textit{port a} is equal to \textit{x}'', whereupon it is automatically determined whether \textit{x} should be driven, or observed and compared. Using the same adder concept described before, but combining them into a single Stream and port with a Reversed child Stream to indicate a response, can be represented as follows:

\begin{lstlisting}[basicstyle=\ttfamily\scriptsize]
adder.add = {
  in1: ("01", "01", "10"),
  in2: ("01", "00", "01"),
  out: ("10", "01", "11"),
};
\end{lstlisting}

Finally, while transactions on ports are not necessarily interdependent, it is reasonable to expect that they will be in many cases. While stateless behavior can be tested in parallel, as each transfer still requires a valid handshake, components which do observe state require that transactions on ports can be asserted in a specific sequence: A counter which accumulates based on input transfers and always drives its output with its current value, or an instruction for a state machine, require that the transfer on the input succeeds before the value on the output is tested. To this end, our design for the testing grammar also includes \textit{sequences} of explicit stages; the assertions within each stage still happen in parallel, but each stage must successfully pass before the assertions in the next stage are performed:

\begin{lstlisting}[basicstyle=\ttfamily\scriptsize]
sequence "sequence name" {
  "initial state": {
    counter.count = "0000";
  }, "increment": {
    counter.increment = "1";
  }, "result state": {
    counter.count = "0001";
  },
};
\end{lstlisting}

\subsection{Substitutions}\label{subsec:substitution}

While the transaction-level verification above can address many kinds of behavior, it cannot account for instances in which inputs and outputs are to be determined at runtime, when a Stream features a \textit{user} signal, or when the behavior of another dependency cannot be simulated for the purposes of an assertion. These cases can instead make use of the IR's ability to quickly compose top-level designs, provide multiple implementations for the same Interface, and subset Streamlets into Interfaces, as described in Section \ref{sec:component_implementation}.

When a dependency cannot be simulated, because it depends on specific hardware, for example, or when it has not been implemented yet, it can be \textit{substituted} with a stub or mock Streamlet. This way, the Streamlet under test can be verified independently. Such substitutes can also be useful to support more complex test cases, by creating components which generate inputs and/or verify outputs. As a simple example, a random number generator component could be paired with a known-good, software-based adder to verify the results of an adder hardware design.

Initially, these substitute components and designs should be separated from the backend's ``proper'' output through namespaces, though we are actively considering making substitutions of Streamlet instances in structural implementations a part of the IR itself. This way, the IR and backend can ensure such explicit substitutions are only used for testing.



\section{Implementation}\label{sec:implementation}
In order to demonstrate the intermediate representation's capabilities and evaluate various approaches, we implemented a prototype toolchain\footnote{\url{https://github.com/matthijsr/til-vhdl/tree/main}}. This toolchain consists of a query system for storing and retrieving the IR's declarations and expressions on-demand, a preliminary grammar and parser which stores its results in the query system, and a backend which uses the query system and emits VHDL.


\subsection{Query System}\label{subsec:query}

The first component of the prototype toolchain is the query system for storing and computing information of the IR. The decision to use a query system rather than more traditional passes of compilation was inspired by work on the Rust compiler \cite{rustcompilerteamqueries2021} and implemented using the Salsa framework \cite{salsa-rssalsa2022}. The advantage of such a system is that information can be retrieved or computed on-demand, and the results of previously executed queries are automatically stored, and only re-computed when their dependencies change.



The query system's database stores type, Interface, Streamlet, Implementation and Namespace declarations. The primary output of the system as a whole is a simple ``all streamlets'' query, which returns all Streamlet declarations from a given input Project. Afterwards, a backend can use other queries, such as a query for splitting a Stream into physical streams, for computing further details as needed.

Another use-case for the query system is the high-level assertions described in Section \ref{subsec:assertions}; converting abstract streams of data on a logical Stream into appropriate, generic calls to the signals that make up its physical streams. Through these functions, a backend would only need to implement the methods for addressing physical streams in order to support these complex, abstract assertions. These are still a work in progress, however, as only methods for transfers on physical streams have been implemented thus far, and the means for declaring and converting transfers on logical Streams to these physical stream transfers have yet to be fully realized.


\subsection{Grammar and Parser}

While the query system is effectively an implementation of the IR in its own right, text-based representations are more portable and can allow for more flexible expressions. Furthermore, a purpose-built language reduces the amount of scaffolding required when testing complete projects in the IR, as compared to setting up the query system manually.

To this end, our prototype toolchain also features a simple grammar (referred to as Tydi Intermediate Language, or \textit{TIL}) and parser, implemented using Chumsky \cite{barrettochumsky2022}. Using the parser, a project expressed in TIL can be stored in the query system. TIL also served as a more stable target for a front-end, computation-oriented language (called Tydi-lang) which was being developed in parallel with the IR, as mentioned in Section \ref{sec:related}.


TIL features expressions for declaring namespaces, types, Interfaces, Streamlets and Streamlet implementations, as well as some syntax sugar for subsetting Streamlets into interfaces. This grammar has been fully implemented in the prototype toolchain, in that it can also be emitted to VHDL using the backend described in the next subsection.

\textbf{Namespaces} are simple containers for other declarations, their only innate property is their name, which can be expressed as a \textit{path}. Note that paths in this context are purely abstract, and do not reflect any hierarchy in the grammar or IR itself, they can simply be used to \textit{communicate} hierarchy to a backend, and/or propagate it from a front-end.

\includegraphics[width=0.5\linewidth]{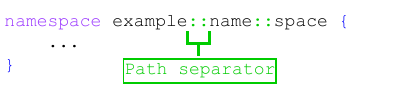}

The \textbf{types} described in Section \ref{subsec:types} can be declared using the \textit{type} keyword, an identifier, and an expression. Type expressions either reference these identifiers, or directly describe the type's properties.

\includegraphics[width=0.6\linewidth]{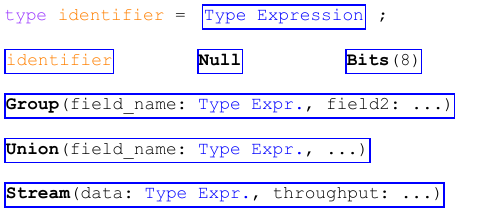}

\textbf{Interfaces}, as described in Section \ref{subsec:interfaces} are collections of ports and (clock and reset) domains. They can be separately declared with an identifier, to enable reuse.


\includegraphics[width=0.8\linewidth]{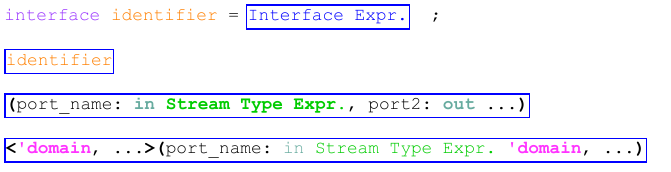}

There are two kinds of \textbf{implementations}, \textit{links} to behavior, and \textit{structural} implementations which connect Streamlets declared in the IR. This is elaborated on in Section \ref{sec:component_implementation}. Links simply use double-quotes to enclose a path to a directory, while structural implementations features two expressions: One to create a Streamlet \textit{instance} and connect the Interface's domains, and another to connect ports between instances and/or the enclosing Streamlet.


\includegraphics[width=0.8\linewidth]{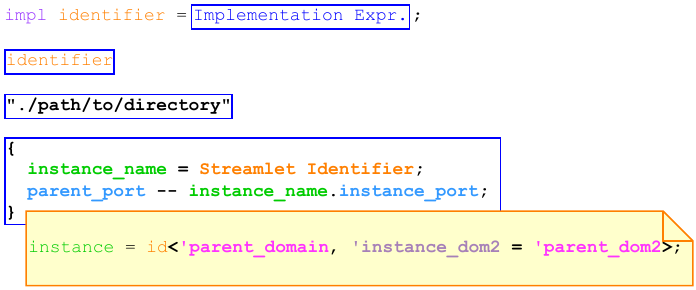}

\textbf{Streamlets} are a combination of the expressions above, and consist of an Interface and optionally an implementation. These are intended to be the output of a backend.

\includegraphics[width=0.75\linewidth]{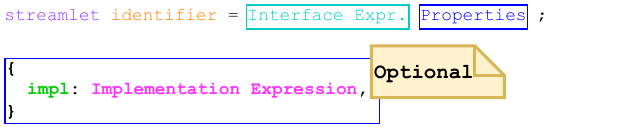}

Finally, \textbf{Documentation} is expressed by enclosing text with \textit{\#} signs, and must precede their subject, as shown in Listing \ref{lst:docex}. As explained in Section \ref{subsec:interfaces}, documentation is distinct from comments in that it is an actual property of Streamlets, ports, and implementations.

\begin{lstlisting}[basicstyle=\ttfamily\scriptsize,caption={Documentation Example},label={lst:docex}]
#documentation (optional)#
streamlet comp1 = (
    // This is a comment
    a: in stream,
    b: out stream,
    #this is port
documentation#
    c: in stream2,
    d: out stream2,
);
\end{lstlisting}

For a complete example, see: \url{https://github.com/matthijsr/til-vhdl/tree/main/demo-cmd/til_samples/paper_example}

\subsection{VHDL Backend}
In order to verify that the IR could actually be compiled to a hardware description, we include a VHDL backend as part of the prototype. As all concepts expressed in the IR would need to be emitted to VHDL, this allowed us to explore which properties were necessary or helpful for targeting hardware.

VHDL was chosen as the target because it is well-supported by multiple toolchains for both synthesis and simulation, and simply because its syntax was most familiar to us. Similar methods as those for emitting VHDL can be employed when emitting other hardware description languages, such as Verilog, FIRRTL and LLHD.

The ``passes'' used when emitting to VHDL in this example backend are intentionally very simple (for instance, while namespaces could correspond to their own VHDL packages, all namespaces are instead combined into a single package), though they do leverage the the query system's ability to incrementally compute and retrieve information:
\begin{enumerate}
    \item The ``all streamlets'' query described in section \ref{subsec:query} is used to retrieve all the Streamlet declarations in the project.
    \item For each Streamlet, the Streams that make up its Interface are split into physical streams, of which the signals are converted into ports. These ports make up a component with a unique name based on the Streamlet declaration and the namespace in which it was declared. These components are added to a single VHDL package.
    \item For each Streamlet, an architecture declaration is either imported or generated:
    \begin{enumerate}
        \item Streamlets without an implementation simply result in an empty architecture.
        \item Streamlets with a linked implementation are imported by looking for an appropriately named \textit{.vhd} file at the given location, an empty architecture is generated at the location if no such file exists.
        \item Streamlets with a structural implementations result in a generated architecture in which port mappings represent Streamlet instances, and signals are used to connect the appropriate ports between instances and the enclosing Streamlet.
    \end{enumerate}
\end{enumerate}

When emitting VHDL, \textit{documentation} from the IR is converted into comments, as shown in  Listing \ref{lst:vhdldoc}. 

\begin{lstlisting}[language=VHDL,basicstyle=\ttfamily\scriptsize,caption={Documentation from Listing \ref{lst:docex} propagating to VHDL},label={lst:vhdldoc}]
-- documentation (optional)
component my__example__space__comp1_com
  port (
    clk : in std_logic;
    rst : in std_logic;
    a_valid : in std_logic;
    a_ready : out std_logic;
    a_data : in std_logic_vector(53 downto 0);
    b_valid : out std_logic;
    b_ready : in std_logic;
    b_data : out std_logic_vector(53 downto 0);
    -- this is port
    -- documentation
    c_valid : in std_logic;
    c_ready : out std_logic;
    c_data : in std_logic_vector(53 downto 0);
    d_valid : out std_logic;
    d_ready : in std_logic;
    d_data : out std_logic_vector(53 downto 0)
  );
end component;
\end{lstlisting}


\section{Evaluation}\label{sec:evaluation}
The prototype toolchain was developed not just as a demonstration, but also to test different approaches and verify their effectiveness. This section contains our evaluation of the prototype toolchain and its features.

\subsection{Tydi Specification}



As a result of explicitly translating the Tydi specification to code, a few oversights and contradictions in the specification came to light. These issues have since been communicated to people working on the Tydi specification, and are being resolved. To accelerate work on the prototype, a few proposed changes have been employed. What follows is a selection of these issues (a), and their proposed solutions (b):

\begin{enumerate}
    \item Directly nested Streams which must both be retained:
    \begin{enumerate}
        \item If a Stream has a direct child Stream (as its \textit{data}), and both have a \textit{user} signal and/or \textit{keep} enabled, it is impossible to create uniquely named physical streams for both.
        \item The prototype toolchain simply returns an error when such an event occurs.
    \end{enumerate}
    \item Strobe assertions and start/end indices may conflict:
    \begin{enumerate}
        \item Physical streams at higher complexities will have strobe, and start/end index signals, to ensure compatibility with lower complexities. It is not specified which signals are significant: indices may indicate a different range of lanes is active than the strobe signal does.
        \item For our work on transfer-level verification, we assume that the start and end indices are only significant when \textit{all} strobe bits are asserted active.
    \end{enumerate}
    \item Element lanes cannot be marked inactive at lower complexities when \textit{dimensionality} is 0:
    \begin{enumerate}
        \item Tydi's documentation on ``signal omission'' \cite{vanstratenphysical2021} suggests that the \textit{end index} signal is contingent on \textit{complexity} $\geq5$ or \textit{dimensionality} $>0$ and \textit{throughput} $>1$. This would result in Streams with multiple element lanes but no dimensionality and \textit{complexity} $<5$ being incapable of disabling element lanes.
        \item The toolchain assumes the \textit{end index} signal is solely contingent on \textit{throughput} $>1$.
    \end{enumerate}
\end{enumerate}

\subsection{Readability}\label{subsec:readability}


As the IR relies on other languages to express functionality, it will generally be necessary for the descriptions a backend \textit{does} generate to be readable by designers, barring a frontend emitting both the IR and the behavioral descriptions. To this end, our IR exposes ``documentation'' to backends, enabling designers to propagate some intent to component templates and interfaces. Our prototype VHDL backend propagates this documentation as comments, and generates indented VHDL with port and signal names derived from the TIL port and field names.

There is one area in which much information and readability is lost, however: The physical streams emitted by the VHDL backend feature standard \textit{data} and \textit{user} signals as bit vectors, meaning that the names of element fields of Groups and Unions are lost. As described in Section \ref{subsec:intrinsics}, the Tydi documentation describes alternative ways to represent physical streams to retain this information. For instance, Groups and Unions could be expressed as record types in VHDL, multiple element lanes as arrays of the base type, and even physical streams themselves could be collected into records (split into separate records for up and downstream signals). These are not only useful for implementation, but can also provide more information when simulating a design.

In fact, the \textit{Implementations} section of the original Tydi paper \cite{peltenburgtydi2020} assumes that designers would prefer such a solution, and illustrates that automatically generating such records from Tydi logical types would greatly reduce the number of lines of code designers would need to write. To better enable such alternative representations, we are considering making changes to the IR to require type identifiers, rather than storing only the official properties of logical types as in Section \ref{subsec:interface_compat}. Doing so would allow a backend to generate alternative representations with meaningful type names, which could then be directly reused by multiple interfaces, albeit at the expense of being able to directly connect physically compatible types.




\subsection{Hardware Description Effort}




The goal of the IR is to describe streams carrying complex data structures more effectively than conventional HDLs. As such, while ``lines of code'' is not an especially relevant metric for an IR overall, it can be applied to the amount of effort required to express interfaces and connections. To evaluate the IR's effectiveness in this regard, we declared Tydi equivalents of the AXI4-Stream \cite{armlimitedamba2010} and AXI4 \cite{armlimitedintroduction2021} interface standards. The AXI4-Stream equivalent and the resulting (VHDL) signals are shown in Listings \ref{lst:axi4streamtil} and \ref{lst:axi4streamvhdl}, while AXI4 was spread over 5 Streams for Address Write, Write Data, Write Response, Address Read, and Read Data.

\begin{lstlisting}[basicstyle=\ttfamily\scriptsize,caption={An AXI4-Stream-equivalent interface in TIL.},label={lst:axi4streamtil},escapechar=@]
    type axi4stream = Stream (
        data: Union (
            data: Bits(8),
            null: Null, // Equivalent to TSTRB
        ),
        throughput: 128.0, // Data bus width
        dimensionality: 1, // Equivalent to TLAST
        synchronicity: Sync,
        complexity: 7, // Tydi's strobe is equivalent to TKEEP
        user: Group (
            TID: Bits(8),
            TDEST: Bits(4),
            TUSER: Bits(1),
        ),
    );

    @\textcolor{gray}{streamlet example = (}@
        axi4stream: in axi4stream,
\end{lstlisting}

\begin{lstlisting}[basicstyle=\ttfamily\scriptsize,caption={Result of Listing \ref{lst:axi4streamtil} in VHDL.},label={lst:axi4streamvhdl}]
axi4stream_valid : in std_logic;
axi4stream_ready : out std_logic;
axi4stream_data : in std_logic_vector(1151 downto 0);
axi4stream_last : in std_logic;
axi4stream_stai : in std_logic_vector(6 downto 0);
axi4stream_endi : in std_logic_vector(6 downto 0);
axi4stream_strb : in std_logic_vector(127 downto 0);
axi4stream_user : in std_logic_vector(12 downto 0);
\end{lstlisting}

Once a Stream type has been declared, it can be easily reused for any number of ports, and ports only require one expression (\lstinline{port_a -- port_b;}) to connect, which is far fewer than the signals which make up a stream (or AXI4 channel). Table \ref{tab:loc_vs_signals} illustrates this difference: The AXI4-Stream equivalent requires a single Stream overall, while AXI4 requires a Stream per channel, and can be either split across multiple ports, or combined into a Group with Reverse Streams for the Read Data and Response channels, depending on the use case. Both result in identical physical streams, but using multiple ports allows for them to be connected to different Streamlets if necessary. The table shows that the number of lines of code for a VHDL AXI4 equivalent representation is 28 compared to only a single line of code for TIL. In the same way, for AXI4 streams, 8 lines of code are required in VHDL compared to only 1 in TIL.

\begin{table}[h]
\centering
\begin{tabular}{l|ll}
                          & Type Declaration & Interface \\ \hline
AXI4 equiv. (TIL)         & 48*              & 5         \\
AXI4 equiv. (TIL, Group)  & 59*              & 1         \\
AXI4 equiv. (VHDL)        & -                & 28        \\
AXI4                      & -                & 44        \\ \hline
AXI4-Stream equiv. (TIL)  & 15*              & 1         \\
AXI4-Stream equiv. (VHDL) & -                & 8         \\
AXI4-Stream               & -                & 9        
\end{tabular}
\caption{Lines of code to represent an interface in TIL, compared to the resulting number of signals in VHDL or for an equivalent interface standard. *Only required once.}
\label{tab:loc_vs_signals}
\end{table}

\section{Conclusion}\label{sec:conclusion}

This paper presents an IR for defining interfaces and integrating components using the Tydi specification. The Tydi-IR can be used to express how composite and
variable-length data structures are transferred over streams using
clear, data-centric types. These data types are extensively used in
many application domains, such as big data and SQL applications. 
The IR prototype toolchain used in this paper features the ability to efficiently express Tydi interfaces and connect components using a simple grammar, and emit these as VHDL components and architectures. We emphasize the ability to propagate high-level, abstract properties down from the IR (and any potential front-end) to the target language, to improve readability and more easily verify its outputs. As a result, we determined that emitting alternative representations for Tydi's interfaces to retain type information could improve readability further, and identified potential changes to the IR to better enable this. The prototype toolchain also features initial work on high-level transaction-level verification to specify intended behavior, and a proposed syntax for such tests. Examples show that while our IR requires only a single line of code to represent stream interfaces, an equivalent VHDL representation would require anywhere from 8 to 28 lines of code to describe the same interface.



\bibliographystyle{ACM-Reference-Format}
\bibliography{bib/Thesis}

\end{document}